\documentclass[prb,aps,nobibnotes,twocolumn]{revtex4}%endfloats
\usepackage{graphicx}%
\usepackage{dcolumn}
\usepackage{amsmath}   
\usepackage{bm}
\usepackage{dcolumn}
\usepackage{longtable}
\usepackage{pifont,color}
\begin{document}

\title{\centering\Large\bf What are the benefits of bound (protonation) states
                           for the electron-transfer kinetics? } 
\author{Dmitry V.\ Matyushov}
\affiliation{Center for Biological Physics, 
  Arizona State University, PO Box 871604, Tempe, AZ 85287-1604}
\date{\today}
\begin{abstract}
  We describe a model of electron transfer reactions affected by local
  binding to the donor or acceptor sites of a particle in equilibrium
  with the solution. The statistics of fluctuations of the
  donor-acceptor energy gap caused by binding/unbinding events are
  non-Gaussian, and the resulting free energy surfaces of electron
  transfer are non-parabolic. The band-width of the charge-transfer
  optical transition is predicted to pass through a maximum as a
  function of the concentration of binding particles in the solution.
  The model is used to rationalize recent observations of
  pH-dependence of electron transfer rates involving changes in the
  protonation state of the donor-acceptor complex.
\end{abstract}
\preprint{Submitted to J.\ Phys.\ Chem.\ B}
\maketitle

\section{Introduction}
\label{sec:1}
Many redox reaction in chemistry and biology involve bound states
which are weaker than common chemical bonds, but stronger than
intermolecular interactions in bulk molecular liquids. A prominent
example of such association is binding of water molecules to solutes
via hydrogen bonds. Since the strength of such bonds typically varies
between different electronic states of the solute, water association
can be recorded by optical
solvatochromism.\cite{Reichardt:94,DMjpcb:97,SolvPolarity:04,Agmon:05}
Another example is the one of proton equilibria involved in biological
electron transfer (ET) reactions responsible for photosynthesis and
respiration.\cite{McEvoy:06} In Photosystem II, the primary donor
P$_{680}^+$ is oxidized by tyrosine which changes its pK$_a$-value
from 10 to $-2$ upon oxidation. As a result, it cannot hold onto its
phenolic proton in aqueous solution releasing it in what is argued to
be a concerted electron-proton transfer mechanism.\cite{Cukier:04}
Another related example is photoacidity when photoexcited
intramolecular charge transfer lowers pK$_a$ for the release of a
proton to the solvent.\cite{Agmon:05} In view of the wide spread of
such reactions, often referred to as proton-coupled
ET,\cite{Cukier:96,Soudackov:00,Hammes-Schiffer:01,Cukier:04} in
biological systems in general and enzymetic reactions in
particular,\cite{Ferguson-Miller:96,Kirby:97} one wonders if their
occurrence is just a coincidence caused by ubiquity of acid-base
equilibria in proteins, or, alternatively, protonation/deprotonation
transitions are used by nature to fine-tune the ET energetics. This
question, also used for the paper title, is the subject of this study.

\begin{figure}
  \centering
  \includegraphics*[width=6.5cm]{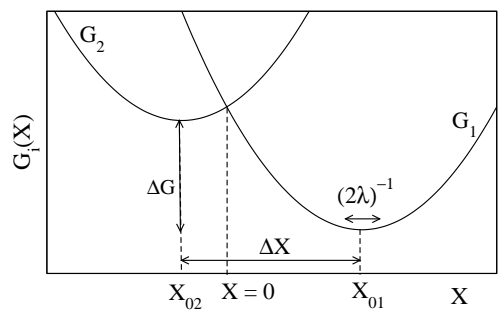}
  \caption{Diagram of the free energy surfaces of ET obtained from
    three parameters, the reaction free energy $\Delta G$, the free energy
    curvature $(2\lambda)^{-1}$, and the Stokes shift (separation between
    the minima $\Delta X$). The reaction coordinate $X$ is the energy
    separation between the donor and acceptor electronic states. The
    activated state is given by the resonance of the donor and
    acceptor levels, $X=0$, while $X_{01}$ and $X_{02}$ indicate the
    positions of the minima. }
\label{fig:1} 
\end{figure}

It has been argued that coupling of ET to the change in protonation
alters the reaction free energy, often turning uphill reactions along
the ET coordinate into downhill reactions along a combined
electron-proton reaction path.\cite{Mayer:04} In order to make this
argument more qualitative, at least one more parameter, the
reorganization energy, needs to be
specified.\cite{Marcus:93,Borgis:96} The picture of crossing parabolas
used to calculate barriers of charge-transfer
reactions\cite{Marcus:65} involves in fact three components: the
reaction free energy $\Delta G$ representing the vertical shift of the
minima, the curvature $(2\lambda)^{-1}$ given in terms of the reorganization
energy $\lambda$, and the horizontal shift of the parabolas' minima
responsible for the Stokes shift, $\Delta X=X_{01}-X_{02}$ (Figure
\ref{fig:1}).

The reaction coordinate $X$, used to monitor the progress of a
charge-transfer reaction, is the instantaneous energy gap between the
donor and acceptor electronic energy levels.\cite{King:90} The
vertical axis in Figure \ref{fig:1} refers to \textit{free energies}
$G_i(X)$ as functions of \textit{energy} $X$. They cross at zero
energy gap $X=0$ allowing electronic tunneling ($i=1,2$ refer to
reactants and products, respectively). In fact, the Marcus-Hush
description of ET\cite{Marcus:89} imposes one more
additional constraint requiring the Stokes shift to be identically
equal to $2\lambda$,
\begin{equation}
  \label{eq:0} 
    \Delta X = 2 \lambda = \beta \langle (\delta X)^2 \rangle 
\end{equation}
where $\beta = 1/(k_{\text{B}}T)$.  Therefore, only two parameters, $\Delta G$
and $\lambda$, are required to determine the activation barrier.  The strong
link between the Stokes shift and the reorganization energy is a
consequence of the Gaussian statistics of $X$.  Once the statistics
become non-Gaussian, the free energy surfaces $G_i(X)$ are
non-parabolic and eq \ref{eq:0} does not hold.\cite{DMacc:07} Here, we
propose a simple model for a charge-transfer reactions triggering the
change in the state of binding of some particle dissolved in the
solvent.  In this development, we are less concerned with the details
of a particular binding mechanisms, but more interested in addressing
a more general question of whether the involvement of bound states can
potentially provide a new mechanism of tuning the energetics of ET not
incorporated in the picture of crossing parabolas. We find that the
statistics of fluctuations of the donor-acceptor energy gap, induced
by local binding/unbinding events, are non-Gaussian thus violating the
link between the Stokes shift and reorganization energy given by eq
\ref{eq:0}. The main consequence of this result is more flexibility,
compared to the standard picture, in tuning the energetics of ET.

\section{Model} 
\label{sec:2}
We will consider a somewhat simplified situation in which the bound
state is fully thermalized on the time-scale of ET i.e.\ on the time
$\tau_{\text{ET}}=k_{\text{ET}}^{-1}$ required for the system to climb
the top of the potential barrier from the equilibrium bottom of the
free energy surface ($k_{\text{ET}}$ is the ET rate constant). This
assumption implies that populations of bound and unbound states follow
Boltzmann statistics along the ET reaction coordinate, a situation
similar to the treatment of intramolecular vibrations in Sumi-Marcus
model.\cite{Sumi:86} This condition can be achieved when the rate of
release of the binding particle B, $k_b$, is much higher that the rate
of ET
\begin{equation} 
\label{eq:1} 
k_{\text{ET}} \ll k_b 
\end{equation}

When this requirement is satisfied, one can use statistical mechanics
to construct a one-dimensional free-energy surface for the ET reaction
in which the exchange between localized and delocalized states of the
binding particle produce fluctuations of the donor-acceptor energy
gap, just as any other solvent mode interacting with the transferred
electron.\cite{Stuchebrukhov:03}

\begin{figure}
  \centering
  \includegraphics*[width=7cm]{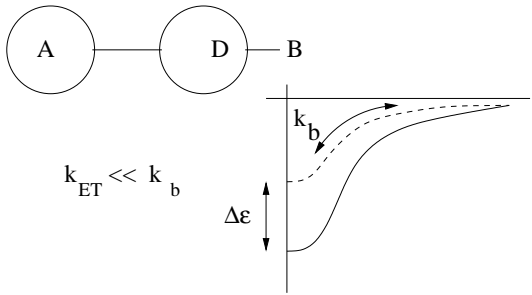}
  \caption{Diagram of particle B exchanging between the bound and
    dissolved states with the rate constant $k_b$. The equilibrium
    population of the bound state assumes that the unbinding rate is
    much faster than the rate of ET ($k_{\text{ET}}\ll k_b$); $\Delta
    \epsilon=\epsilon_2-\epsilon_1$ denotes the change in the binding energy caused by the
    electronic transition. }
  \label{fig:2}
\end{figure}

The release of one particle from the bound localized state requires
the Gibbs energy $g_i$ balancing the binding energy $\epsilon_i$ with an
entropy gain from moving the particle into the bulk. The binding
energy $\epsilon_i$ is generally different in the initial, D--A ($i=1$), and
final, D$^+$--A$^-$ ($i=2$), configurations of the donor-acceptor
complex (Fig.\ \ref{fig:2}). The difference in free energies needed to
release particle B to the solution is equal to the difference in
binding energies, $\Delta g = \Delta\epsilon = \epsilon_2-\epsilon_1$. Therefore, this quantity can
be estimated from the difference of the equilibrium dissociation
constants pK$_a^{(i)}$:
\begin{equation}
 \label{eq:2}
  \Delta \epsilon  = k_{\text{B}}T  \ln 10 \Delta \mathrm{pK}_a
\end{equation}

The requirement of non-adiabaticity of ET, which we implicitly assume
here, imposes an additional constraint on the rate of bound particle
release.  The change of the donor and acceptor energy levels caused by
the motion of the bound particle should not break the Landau-Zener
non-adiabaticity condition close to donor-acceptor energy resonance:
\begin{equation}
 \label{eq:3}
  \frac{2\pi}{\hbar} |V_{DA}|^2/ \dot\epsilon \ll 1
\end{equation}
Here, $V_{DA}$ is the donor-acceptor electronic coupling. Assuming the
rate of binding energy change as $\dot \epsilon \simeq k_b \Delta \epsilon $ and $V_{DA}$ of
the order of 1 cm$^{-1}$, eq \ref{eq:3} requires the rate of release
of B to be faster than 1 ns$^{-1}$.  This threshold rate is comparable
to the rate of proton release from bound protonation states of
proteins,\cite{Gutman:97} while the time of water exchange between the
protein surface and the bulk is faster, in the range of tens of
picoseconds.\cite{Halle:04}

Once particle B is released from its bound state, it becomes a part of
the bulk, which is the aqueous solution or a solvent mixture for water
binding or the ionic atmosphere for ionic (proton) association. The
interaction of the bulk with electronic states of the donor and
acceptor is relatively well understood. For water binding,
fluctuations of the dipolar polarization produce the thermal noise.
For electrolytes, the Debye-H{\"u}ckel electric field both shifts the
donor-acceptor energy gap and creates its fluctuations thus producing
a corresponding reorganization energy.\cite{German:92} This ionic
reorganization energy $\simeq \kappa^2 e^2 R / \epsilon_s$ scales quadratically with
the inverse Debye-H{\"u}ckel length $\kappa$ and linearly with a solute
dimension $R$.  Since it is also inversely proportional to the static
dielectric constant of the solvent $\epsilon_s$, the effect of fluctuations
of the ionic atmosphere in polar solvents is commonly small relative
to the reorganization energy of polarization fluctuations. We will
therefore assume that once particle B is released from its bound
state, it does not interact any more with the transferred electron and
becomes a part of the many-particle electrostatic potential.

This approximation allows us to apply the tight-binding approximation
and to use the following expression for the Hamiltonian of the
donor-acceptor system in a polar solvent
\begin{equation}
  \label{eq:4}
  \begin{split}
  H_i & = E_{0i} - \left(\mathbf{E}_i^{u} |u\rangle\langle u| + \mathbf{E}_i^{b} |b\rangle\langle b|\right)*\mathbf{P} \\
    & - \epsilon_i |b\rangle \langle b|     + \frac{1}{2} \mathbf{P}*\chi^{-1}*\mathbf{P} 
  \end{split}
\end{equation}
Here, $E_{0i}$ incorporates the vacuum energy and the free energy of
solvation by the electronic degrees of freedom of the solvent and
$\mathbf{E}_i^{u,b}$ denotes the vacuum electric field of the
donor-acceptor complex in two ET states ($i=1,2$) and two binding
states (u,b).  Also, $|b\rangle \langle b|$ and $|u\rangle \langle u|$ in eq \ref{eq:4}
describe the population operators for the bound (b) and unbound (u)
states of particle B with the binding energy $\epsilon_i$.  One can expect
only a small change in the electric field of the donor-acceptor
complex for weak binding of a neutral molecule, $\mathbf{E}_i^b \simeq
\mathbf{E}_i^u$. However, the dependence of the electric field on the
binding state needs to be incorporated for equilibria of charged
particles.

In application to protonation/deprotonation equilibria, the
Hamiltonian considered here (eq \ref{eq:4}) is different from the ones
used by Cukier\cite{Cukier:96} and Soudackov and
Hammes-Schiffer.\cite{Soudackov:00} In their case, the proton was
considered to move between two localized states within the
donor-acceptor complex, while in the problem considered here the
proton is released to the bulk and loses any connection to electronic
transitions within the donor-acceptor pair except for the influence of
the Debye-H{\"u}ckel field it becomes a part of. The different physics
of the problem considered here demands the different Hamiltonian.
 
The electric field interacts with the (nuclear) dipolar polarization
of the solvent $\mathbf{P}$, and the asterisk in eq \ref{eq:4} implies
both the tensor contraction and the volume integration.  The
statistics of $\mathbf{P}$ are Gaussian with the response function
$\mathbf{\chi}$ such that the solvent reorganization energy is
\begin{equation}
  \label{eq:5}
  \lambda_{u,b} = \frac{1}{2}\mathbf{\Delta E}_{u,b}*\chi*\mathbf{\Delta E}_{u,b}, \quad 
                 \Delta \mathbf{E}_{u,b}= \mathbf{E}_2^{u,b} -\mathbf{E}_1^{u,b} 
\end{equation}
Although $\lambda_{u,b}$ in eq \ref{eq:5} are formally the reorganization
energies of the dipolar polarization field of the solvent, we will
attach a more general meaning to them as the reorganization energies
of the nuclear degrees of freedom of the bulk. The corresponding
formal definition is easy to obtain from eq \ref{eq:4} by replacing
the Gaussian polarization noise with a sum of statistically
independent Gaussian fields.

The free energies of ET are obtained from the Hamiltonian in eq
\ref{eq:4} by projecting all possible nuclear motions on the reaction
coordinate $X=\Delta H = H_2 - H_1$:
\begin{equation}
  \label{eq:6}
  e^{-\beta G_i(X)} = \int \mathcal{D} \mathbf{P} 
                             \mathrm{Tr}\left[\delta(X - \Delta H) e^{-\beta H_i}\right]
\end{equation}
Here, $\mathcal{D} \mathbf{P}$ implies functional integration over the
field $\mathbf{P}$ and ``Tr'' refers to the sum over bound, $|b\rangle $, and unbound,
$|u\rangle$, states of particle B with statistical weights $f_{b,u}$:
\begin{equation}
  \label{eq:7}
  \mathrm{Tr}[\dots ] = \sum_{a=b,u} \langle a|f_a \dots |a\rangle .    
\end{equation}
The ratio of the statistical weights gives the entropy of releasing
particle B to the bulk, $k_{\text{B}}\ln(f_u/f_b)$.

Taking the integral and trace in eq \ref{eq:6} results in the following
equation for the free energy surfaces of ET
\begin{equation}
  \label{eq:8}
  \begin{split}
  G_i(X)& = G_i^u + \frac{(X-\Delta E_i^u)^2}{4\lambda_u} \\
      & - \beta^{-1}\ln\left(1 + 10^{\mathrm{pK}_a^{(i)}-\mathrm{pB}}f_i(X)\right) 
  \end{split}
\end{equation}
In eq \ref{eq:8}, $G_i^u$ is the free energy of the donor-acceptor
complex in the unbound state. The solvation free energy entering
$G_i^u$ is thus of electrostatic origin and does not include the free
energy of binding particle B.  Correspondingly, the vertical energy
gap $\Delta E_i^u$ in the second summand is given as $\lambda_u + \Delta G_u$ for the
forward transition $1\to 2$ and as $\lambda_u - \Delta G_u$ for the backward
transition $2\to 1$. Binding of particle B is expressed by the term
under the logarithm where function $f_i(X)$ is the ratio of Boltzmann
factors for activation through bound and unbound states:
\begin{equation}
  \label{eq:8-1}
  f_i(X) = \exp\left[-\beta \frac{(X+\Delta \epsilon - \Delta E_i^b)^2}{4\lambda_b} + \beta \frac{(X- \Delta E_i^u)^2}{4\lambda_u}   \right]
\end{equation}
Here, $\Delta E_i^b$ are the vertical ET gaps in the bound state of 
particle B which are given as $\lambda_b + \Delta G_b$ ($i=1$) and $\lambda_b - \Delta G_b$
($i=2$).

In writing eq \ref{eq:8} we have also represented the Boltzmann factor
for the release of particle B in terms of pK$_a^{(i)}$ and pB$= -
\log[B]$ as follows
\begin{equation}
  \label{eq:9}
  e^{-\beta g_i - \beta \Delta G_i }= 10^{\mathrm{pB} - \mathrm{pK}_a^{(i)}}
\end{equation}
In this equation $g_i=\epsilon_i - k_{\text{B}}T\ln(f_u/f_b)$ is the free
energy of releasing the bound particle from the molecular fragment in
the donor-acceptor complex (e.g.\ tyrosine) and $\Delta G_i = G_i^u -
G_i^b$ is the change in the solvation energy of the entire
donor-acceptor complex caused by unbinding event. The equilibrium
constants $\mathrm{pK}_a^{(i)}$ thus reflects the equilibrium of the
entire complex and can be modified compared to the equilibrium
constants of the molecular fragment alone.

The free energies $G_i(X)$ can be used to calculate the first and
second moments of the reaction coordinate $X$, e.g.
\begin{equation}
  \label{eq:10}
  \langle X\rangle_i = Q_i^{-1} \int X e^{-\beta G_i(X)} dX
\end{equation}
where
\begin{equation}
  \label{eq:11}
  Q_i = \int e^{-\beta G_i(X)} dX 
\end{equation}
One gets for the average
\begin{equation}
  \label{eq:12}
   \langle X\rangle_i = (1 - n_i) \Delta E_i^u + n_i (\Delta E_i^b- \Delta\epsilon)
\end{equation}
and for the variance
\begin{equation}
  \label{eq:13}
   \begin{split}
   \langle (\delta X)^2 \rangle_i &=  2k_{\text{B}}T\left[\lambda_b n_i + \lambda_u(1-n_i)\right]\\
            &  +n_i(1-n_i)\left[\Delta E_i^u - \Delta E_i^b + \Delta \epsilon\right]^2
   \end{split}
\end{equation}
In eqs \ref{eq:12} and \ref{eq:13},
\begin{equation}
  \label{eq:14}
    n_i = \left[1 + 10^{\mathrm{pB}-\mathrm{pK}_a^{(i)}} \right]^{-1} 
\end{equation}
is the equilibrium population of the bound site. 

\begin{figure}
  \centering
  \includegraphics*[width=6.5cm]{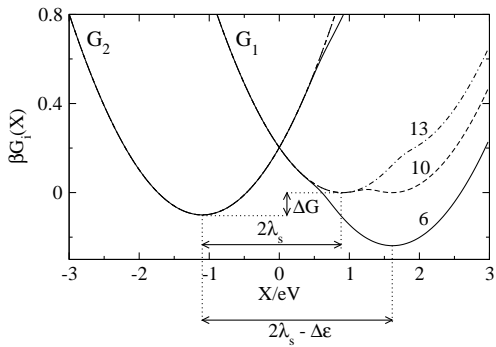}
  \caption{Free energy surfaces of ET (eq \ref{eq:14-1}) at different
    values of pB indicated in the plot assuming pK$_a^{(1)}=10$ and
    pK$_a^{(2)}=-2$ (equilibrium constants of tyrosine). The final
    charge-transfer state ($i=2$) is always deprotonated while the
    initial charge-transfer state ($i=1$) is protonated at
    pB$<$pK$_a^{(1)}$ and is deprotonated at pB$>$pK$_a^{(1)}$. The
    Stokes shift for the protonated state 1 is $2\lambda_s - \Delta\epsilon = 2.71$ eV
    and is equal to $2\lambda_s=2$ eV for the deprotonated state 1. The
    solvent bulk component of the reaction Gibbs energy reaction is $\Delta
    G = -0.1$ eV. }
  \label{fig:3}
\end{figure}

Equations \ref{eq:6}--\ref{eq:14} provide a general description of ET
when binding affects both the statistics of the energy gap
fluctuations and the electronic density responsible for the electric
field.  In order to reduce the number of independent parameters, we
first neglect the second effect assuming that the electric field does
not change with binding, i.e.\ $\lambda_u=\lambda_b=\lambda_s$, $\Delta E_i^u=\Delta E_i^b=\Delta E_i$,
and $G_i^u=G_i^b=G_i$. This situation most closely corresponds to
binding of a neutral molecule (water) while the case of a charged
particle (protonation) is postponed to the next section.

The free energy surfaces of ET can then be written as follows:
\begin{equation}
  \label{eq:14-1}
  \begin{split}
  &G_i(X) = G_i + \frac{(X-\Delta E_i+\Delta \epsilon)^2}{4\lambda_s} \\
  & - \beta^{-1} \ln\left(10^{\mathrm{pK}_a^{(i)} - \mathrm{pB}} + \exp\left[\frac{\beta\Delta\epsilon}{2\lambda_s}(X-\Delta E_i +\Delta\epsilon/2)\right]\right)
  \end{split}
\end{equation}
where $G_i$ refers to the free energy of the donor acceptor complex
with particle B released to the solution.  

Figure \ref{fig:3} illustrates the change of the free energy surfaces
with pB according to eq \ref{eq:14-1} (pK$_a^{(i)}$ values of
tyrosine). For the reaction $1\to 2$, the system needs to climb the
activation barrier from the bottom of the free energy well $G_1(X)$ to
the crossing point at $X=0$. The bottom of the potential well rises
with increasing pB at pB$<$pK$_a^{(1)}$ thus resulting in a smaller
activation barrier.  The reaction rate depends on pB. The barrier
height stops changing once pB reaches pK$_a^{(1)}$, and the reaction
rate is insensitive to pB at pB$>$pK$_a^{(1)}$. Notice that the rate
of the reverse transition $2\to 1$ remains unchanged in the whole range
of pB values.  This invariance makes the effect of pB on the ET rate
distinct from the effect of the driving force which alters the
activation barriers for both the forward and backward reactions.

When binding does not affect the electric field of the donor-acceptor
complex the equations for the first and second cumulants (eqs
\ref{eq:12} and \ref{eq:13}) simplify to
\begin{equation}
  \label{eq:15}
   \langle X\rangle_i = \Delta E_i - \Delta\epsilon n_i
\end{equation}
and
\begin{equation}
  \label{eq:16}
  \langle (\delta X)^2 \rangle_i = 2k_{\text{B}}T \lambda_s + \Delta\epsilon^2 n_i(1-n_i)   
\end{equation}
Several important observations follow from eqs \ref{eq:15} and
\ref{eq:16}.  First, the binding/unbinding fluctuations break the
connection (eq \ref{eq:0}) between the reorganization energy from the
free energy curvature
\begin{equation}
  \label{eq:17}
  \lambda_i = (\beta/2) \langle (\delta X)^2 \rangle_i
\end{equation}
and the Stokes shift
\begin{equation}
  \label{eq:18}
  \Delta X = 2\lambda_s + \Delta\epsilon \Delta n,\quad \Delta n = n_2 - n_1 
\end{equation}
This effect is the consequence of the local character of the unbinding
events contrasting with the quasi-macroscopic nature of the bulk
fluctuations resulting in eq \ref{eq:0}.  Second, the reorganization
energy in eqs \ref{eq:16} and \ref{eq:17} arising from binding
fluctuations depends on the state of the donor-acceptor complex, i.e.\
$\lambda_1 \neq \lambda_2$. This condition implies that the free energy surfaces
$G_i(X)$ must be non-parabolic\cite{DMacc:07} to comply with the
energy conservation,\cite{King:90} $G_2(X)=G_1(X)+X$.  The effect of thermalized
bound states on ET cannot therefore be accounted for within the
Marcus-Hush formalism. We also note that the variance of the reaction
coordinate in eq \ref{eq:14} does not follow the
fluctuation-dissipation theorem\cite{Landau5} requiring $ \langle (\delta X)^2\rangle_i
\propto T$ and instead has a more complex temperature dependence arising
from the equilibrium dissociation constant entering the equilibrium
population in eqs \ref{eq:13} and \ref{eq:16}.

\begin{figure}
  \includegraphics*[width=6.5cm]{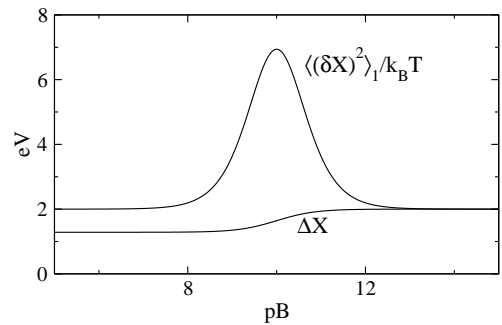}
  \caption{Stokes shift, $\Delta X$, and the reaction coordinate variance,
    $\langle (\delta X)^2\rangle_1$, vs pB for a bound state with the dissociation
    constants of tyrosine (pK$_a^{(1)}=10$ and pK$_a^{(2)}=-2$) and
    the solvent reorganization energy $\lambda_s=1$ eV. }
  \label{fig:4}
\end{figure}

The dependence of both the energy gap variance and the Stokes shift on
pB are shown in Figure \ref{fig:4}. The equality of $\beta \langle (\delta X)^2\rangle_1$
and $\Delta X$, expected from eqs \ref{eq:0}, is seen when both the initial
and final ET states are in the same binding state. They are shifted by
the binding energy $\Delta \epsilon$ when the the binding states are different in
the two ET states (left corner in Figure \ref{fig:4}). The most
interesting region is when pB$\simeq$pK$_a$ and the term proportional to
$n_i(1-n_i)$ in eqs \ref{eq:13} and \ref{eq:16} can potentially
dominate the energy gap variance. Our model thus makes a simple,
experimentally testable prediction that the band-width of a
charge-transfer optical transition (absorption for binding to the
donor) is expected to pass through a maximum as a function of pB.

\section{Protonation affecting electron transfer}
\label{sec:3}
Measurements of oxidation rates of primary pair P$_{680}^+$ by
tyrosine in Photosystem II have produced intriguing
results.\cite{Ahlbrink:98} The reaction rate increases with increasing
pH until it levels off at pH approximately equal to pK$_a$ of phenolic
proton. Similar results were reported by Hammarstr{\"o}m and co-workers
for intramolecular ET from tyrosine to Ru(III) covalently connected in
a donor-acceptor complex.\cite{Sjodin:00,Lomoth:06} In order to
explain the observed pH-dependence, they used the Marcus equation for
the activation free energy in which a pH-dependent redox potential was
substituted. The pH-dependence of the rate arises in their analysis
from the dependence of the tyrosine reduction potential on the
concentration of protons in the solution.  

This practice,\cite{Sjodin:00,Carra:03} which had not been anticipated
in the original Marcus formulation,\cite{Marcus:65} was criticized by
Krishtalik\cite{Krishtalik:03} and, more recently, by Sav{\'e}ant and
co-workers.\cite{Costenin:07} Krishtalik argued that the Gibbs free
energy of ET reactions cannot possibly depend on the ideal mixing
entropy of protons in the bulk, which is the origin of the
pH-dependent term in the Nernst equation for the redox
potential.\cite{Bockris:70} We can only add to this, absolutely
correct, argument that using a pH-dependent $\Delta G$ in the Marcus
equation clearly violates the Franck-Condon principle.  The vertical
\textit{energy} gap $\Delta G + \lambda $ does not involve entropy change since
the nuclei do not move on the time-scale of electronic transitions.
When the entropic pH term appears in $\Delta G$ and is not compensated by a
corresponding term in $\lambda$, the unphysical entropy of protons mixing
appears in the vertical transition energy. The present model allows us
to account for the pH-dependence of the ET rate without introducing
unphysical approximations.

\begin{figure}
  \centering
  \includegraphics*[width=6.5cm]{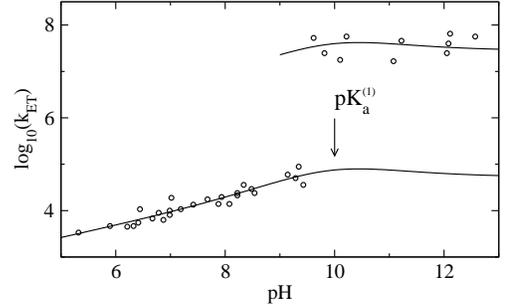}
  \caption{Charge shift rates from tyrosine to Ru(III) from ref
    \onlinecite{Sjodin:00} (points) and the fit of the data using eq
    \ref{eq:14-1} (solid lines).  The lower curve is obtained using
    $\lambda_{u}=2.3$ eV, $\lambda_{b}=1.8$ eV, and the non-ergodicity multiplier
    $\alpha=0.27$ ($\Delta G_u=-0.54$ eV and $\Delta G_b=0.2$ eV are taken from ref
    \onlinecite{Carra:03}).  The upper curve was obtained with the
    same activation parameters by increasing the rate preexponent (see
    text for discussion).  The vertical arrow indicates the pK$_a^{(1)}$
    value for tyrosine at which the ET rate becomes independent of pH.
  }
  \label{fig:5}
\end{figure}

For protonation affecting ET, B$=$H and pB$=$pH. This problem is,
however, potentially more complex than binding/unbinding of neutral
molecules. The process of deprotonation proceeds by formation of a
geminate pair (on the time-scale of picoseconds for photoexcited
states\cite{Agmon:05}) followed by a slower diffusion of the released
proton in the Coulomb potential of the deprotonated negative
charge.\cite{Agmon:05,Perez:07} This slow process may imply that the
unbound proton will not be able to sample all possible states
available to a particle in an ideal solution on the time-scale of ET,
$\tau_{\text{ET}}$.  A full account of such effects\cite{DMacc:07}
presently appears hard to achieve.  We will therefore introduce an
empirical non-ergodicity multiplier $\alpha$ to account for the effects of
insufficient sampling.  The term pK$_a^{(i)}-$pH in eq \ref{eq:8} is
replaced with $\alpha(\mathrm{pK}_a^{(i)}-\mathrm{pH})$, where $\alpha$ is a
non-ergodicity coefficient between zero and unity. With this
correction, the model qualitatively reproduces the dependence of the
ET rate on pH observed by Sj\"odin \textit{et al}.\cite{Sjodin:00} for
the oxidation of tyrosine by Ru(III).
 
Sj\"odin \textit{et al}.\cite{Sjodin:00} have monitored the recovery of
Ru(II) from Ru(III) produced by flash photolysis and accelerated by
electron transfer from tyrosine covalently linked to the ruthenium
complex.  Their observed rates, monitoring the arrival of electrons,
mathematically correspond to projecting the complex, potentially
multi-coordinate,\cite{Agmon:83,Sumi:86,Soudackov:00} dynamics of the
system onto one single ET reaction coordinate, which is exactly the
scenario considered here. Since proton is a charged particle, we need
the full formulation given by eq \ref{eq:8} to analyze the
experimental rates. The reaction Gibbs energies and reorganization
energies of ET will potentially be different for the bound and unbound
ET pathways and indeed the reaction free energies are $\Delta G_u = - 0.54$
eV and $\Delta G_b=0.2$ eV for the unbound and bound proton states,
respectively.\cite{Carra:03} The unknown parameters are the rate
preexponent, reorganization energies $\lambda_{u,b}$, and the non-ergodicity
multiplier $\alpha$.  This latter parameter is expected to be small because
of the slow rate of proton dissociation from tyrosine\cite{Sjodin:00}
and thus a lower extent of modulation of the donor-acceptor gap by
unbinding events than it would be possible for full thermalization.

\begin{figure}
  \centering
  \includegraphics*[width=6.5cm]{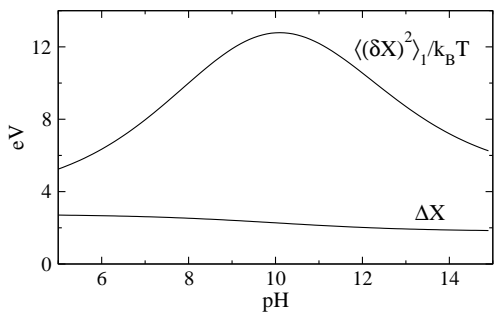}
  \caption{Stokes shift, $\Delta X$, and the variance, $\langle(\delta
    X)^2 \rangle_1$, calculated from eqs \ref{eq:12} and \ref{eq:13} using
    the reorganization parameters obtained by fitting the experimental
    data in Figure \ref{fig:5}.  }
  \label{fig:6}
\end{figure}

In order to see if the model can qualitatively account for the
experimental observations, we have used the rate preexponent,
reorganization energies $\lambda_{u,b}$, and $\alpha$ as free parameters to fit
the data from ref \onlinecite{Sjodin:00}. The result is shown in
Figure \ref{fig:5}.  The fitting curve ($\lambda_u=2.3$ eV, $\lambda_b=1.8$ eV,
and $\alpha=0.26$) of $\log_{10}(k_{\text{ET}})$ vs pH rises linearly with
the slope $\alpha$ when pH is below pK$_a^{(1)}$ and levels off after
reaching this value.  The reorganization energies $\lambda_{u,b}$ here
include reorganization of classical intramolecular vibrations in
addition to solvent reorganization.  The parameters extracted from the
fit of the rates result in a dramatic violation of eq \ref{eq:0} as is
shown in Figure \ref{fig:6}.

Fitting the experimental jump in the rate at
$\mathrm{pH}=\mathrm{pK}_a^{(1)}$ requires a higher value of the rate
preexponent as is shown by the fragment of the curve obtained by using
the same parameters as in the low-pH fit, but varying the preexponent.
The increase of the rate at $\mathrm{pH}=\mathrm{pK}_a^{(1)}$ has been
addressed by Carra \textit{et al}.\cite{Carra:03} The low-pH portion
of the curve corresponds to electronic transition accompanied by
simultaneous proton release (proton-coupled
ET\cite{Hammes-Schiffer:01}). The nonadiabatic matrix element in the
rate preexponent then includes small Franck-Condon overlap between
proton bound and free states.\cite{Cukier:96} This overlap disappears
in the flat portion of the plot at $\mathrm{pH}>\mathrm{pK}_a^{(1)}$
when tyrosine is deprotonated in both ET states. Only electronic
coupling enters the rate preexponent in this regime resulting in a
higher rate.  A more quantitative analysis would require extensive
calculations.  Since the model presented here does not aim at a
detailed quantitative analysis of proton-coupled ET, we limit our
discussion to qualitative observations only.

\section{Conclusions}
\label{sec:4}
Traditional theories of ET in polar liquids emphasize activation of
electronic transitions by long-range, quasi-macroscopic solvent
fluctuations.\cite{Marcus:65} Local binding to the donor and acceptor
sites provides an additional modulation of the donor-acceptor energy
gap. This study addressed the question of whether this additional
thermal noise can be accounted for within the standard framework of
Gaussian fluctuations of the energy gap, that is by adjusting the
magnitudes of the driving force and reorganization energy. 

We have found that the statistics of binding events are non-Gaussian,
and the resultant free energy surfaces cannot be reduced to crossing
parabolas. The model predicts a regime of a significant dependence of
the activation barrier on the concentration of the binding particles
in the solution. When the concentration pB is close to the binding
equilibrium constant pK$_a$, the variance of the energy gap passes
through a maximum which can be observed by spectroscopy of
charge-transfer bands.  Despite these new features, the main effect of
binding is in shifting the free energy gap of ET, as was suggested
previously,\cite{Mayer:04} leaving the reorganization energy mostly
unaffected and within the realm of standard models.  Finally, the
model helps to rationalize some recent observations of the dependence
of rates of intramolecular ET, involving deprotonation of reactants,
on the pH of the solution.

\begin{acknowledgments}
  This work was supported by the NSF (CHE-0616646).
\end{acknowledgments}

%\bibliographystyle{apsrev}
% \bibliography{/home/dmitry/p/bib/chem_abbr,/home/dmitry/p/bib/photosynthNew,/home/dmitry/p/bib/glass,/home/dmitry/p/bib/et,/home/dmitry/p/bib/dm,/home/dmitry/p/bib/pt,/home/dmitry/p/bib/solvation,/home/dmitry/p/bib/heteroet,/home/dmitry/p/bib/bioet,/home/dmitry/p/bib/protein}

\providecommand{\refin}[1]{\\ \textbf{Referenced in:} #1}

\end{document}